\documentstyle[aps,psfig]{revtex}

\input epsf

\begin{document}
\title{Finite size effects on the phase diagram of a binary mixture confined
between competing walls}
\author{M. M\"{u}ller and K. Binder}
\address{Institut f\"{u}r Physik, Johannes Gutenberg Universit\"{a}t Mainz, D-55099\\
Mainz, Staudinger Weg 7, Germany}
\author{and E. V. Albano}
\address{INIFTA, Universidad de la Plata, CC16 Suc. 4, 1900 La Plata, Argentina}
\maketitle

\begin{abstract}
A symmetrical binary mixture AB that exhibits a critical temperature T$_{cb}$
of phase separation into an A-rich and a B-rich phase in the bulk is
considered in a geometry confined between two parallel plates a distance D
apart. It is assumed that one wall preferentially attracts A while the other
wall preferentially attracts B with the same strength (''competing walls'').
In the limit $D\rightarrow \infty $, one then may have a wetting transition
of first order at a temperature $T_{w}$, from which prewetting lines extend
into the one phase region both of the A-rich and the B-rich phase. It is
discussed how this phase diagram gets distorted due to the finiteness of $D$%
: the phase transition at $T_{cb}$ immediately disappears for $D<\infty $
due to finite size rounding, and the phase diagram instead exhibit two
two-phase coexistence regions in a temperature range $%
T_{trip}<T<T_{c1}=T_{c2}$. In the limit $D\rightarrow \infty $ $\ $\ $\
T_{c1},T_{c2}$ become the prewetting critical points and $%
T_{trip}\rightarrow T_{w}$.

For small enough D it may occur that at a tricritical value D$_{t}$ the
temperatures $T_{c1}=T_{c2}$ and $T_{trip}$ merge, and then for $D<D_{t}$
there is a single unmixing critical point as in the bulk but with $T_{c}(D)$
near $T_{w}$. As an example, for the experimentally relevant case of a
polymer mixture a phase diagram with two unmixing critical points is
calculated explicitly from self-consistent field methods.
[will be published in Physica A 279 (No. 1-4) (2000) pp. 188--194.]
\end{abstract}

\section{Introduction}

Although the finite size effects on phase transitions in thin films have
been studied since a long time \cite{1,2,3,4,5,6,7,8,9,10}, only during the
last decade it was discovered \cite{11,12,13,14,15,16,17,18} that in
ferromagnetic Ising films with surface fields of different sign but of the
same strength $\pm H_{1}$ (''competing walls'') novel types of phase
transitions occur: namely, a phase transition occurs for zero bulk field $H$
from a state with an interface freely fluctuating in the center of the thin
film to a state where the interface is bound either to the lower or to the
upper wall confining the film, which then acquires a nonzero (positive or
negative) magnetization. This interface localization-delocalization
transition at $T_{c}(D)$ may be either second \cite{12,13,14,15,16,17} or
first order \cite{13,17,18}, and for film thicknesses $D\rightarrow \infty $
the transition temperature $T_{c}(D)$ does not converge towards the bulk
critical temperature $T_{\text{cb}}$ as usual \cite{1,2,3,4,5,8,9,10}, but
rather towards the wetting transition temperature $T_{\text{w}}(H_{1})$ \cite
{12,13,14,15,16,17,18}.

Now it is well-known that, in general, wetting transitions \cite{19} can be
second or first order \cite{20}. Thus it is plausible that also the
interface localization-delocalization transition can be either second or
first order. However, recently it was shown \cite{17} that also in cases
where the wetting transition is first order, the transition $T_{c}(D)$ may
be second order for small enough thickness $(D<D_{t})$ and become first
order only for $D>D_{t}$. Thus the transition at $T_{c}(D_{t})$ is the
finite-thickness analog of a wetting tricritical point \cite{13,14,15,16,17}.

All this work \cite{11,12,13,14,15,16,17,18} has only considered the case $%
H=0$, however. It is well known of course, that for $D\rightarrow \infty $,
in the case of a first-order wetting transition at $T_{w}(H_{1}<0)$ there
exists at $T=T_{\text{pre}}(H,H_{1})$ a prewetting transition \cite{19,20},
where the distance of the interface bound to the wall discontinuously jumps
from a smaller value to a larger value. While the analog of this prewetting
transition in thin films has been studied occasionally for the case of
''capillary condensation'' (where on both walls fields act that have the
same sign) \cite{7,10}, it is only in the present work that the effect of
prewetting phenomena on the interface localization-delocalization transition
is considered \cite{21}. The physical systems that we have in mind are not
magnets, of course, but rather binary (A,B) mixtures: as is well known, in
an Ising model context the ''magnetization'' simply translates into the
relative concentration $\phi $ if one component (A, say), and the field $H$
translates into the chemical potential difference $\Delta \mu $ between the
species (for simplicity we deal here with perfectly symmetric mixtures for
which the bulk critical concentration is $\phi _{\text{cb}}=0.5$).

However, one important aspect of binary mixtures is that physically it is a
density of an extensive thermodynamic variable (namely $\phi $) that is the
fixed independent thermodynamic variable, rather than the intensive variable 
$\Delta \mu $. As we shall see below, this fact has important consequences
for the phase diagram of confined mixtures: the typical situation is that
one encounters two successive lateral phase separation transitions!

In Section 2 we elaborate these ideas by a qualitative discussion of the
phase diagrams \{both in the space of intensive variables ($\Delta \mu ,T$)
and in the space ($\phi ,T$)\} and of the corresponding physical state of
the confined mixture. Section 3 exemplifies these considerations by
presenting a specific calculation for a symmetric polymer mixture, treated
within a self-consistent field framework \cite{21}.\ There are numerous
experimental studies of confined polymer mixtures \cite{22} and these
systems might be convenient for testing our predictions. Finally section 4
summarizes our conclusions.

\section{Qualitative phase diagrams of confined binary mixtures}

We assume here a binary mixture confined by ''competing'' walls in the sense
that one wall attracts species A with the same strength as the opposite wall
attracts species B, and consider the case of first-order wetting. Then the
topology of the phase diagram can be estimated from the qualitative considerations 
as shown in the left part of Fig.~1: In the
space ($T,\Delta \mu $), bulk ($D\rightarrow \infty $) phase separation
occurs for $\Delta \mu =0$, and the walls are incompletely wet for $T<T_{w}$
but wet for $T_{w}<T<T_{cb}$. From the point $T=T_{w},\Delta \mu =0$ there
extend two (first-order) prewetting lines, which end at the prewetting
critical points $T_{cp}$. These prewetting transitions correspond to
singularities of the surface free energies associated with the lower and
upper wall confining the mixture (Fig.~2). Due to the special symmetries
chosen for our model, both the wetting transitions for the lower and upper
wall coincide, and the prewetting critical temperatures are also the same.
There is a mirror symmetry of the phase diagrams around the line $\Delta \mu
=0$ (upper part) or $\phi =0.5$ (lower part), respectively.

For finite thickness $D$ it may happen, as demonstrated by Monte Carlo
simulations for Ising models with enhanced exchange interactions near the
walls \cite{17}, that a tricritical point at $D=D_{t}$, $T_{c}(D_{t})$, $%
\Delta \mu =0$ (and $\phi =0.5$) occurs. For $D<D_{t}$ then there exists a
single critical point at $T=T_{c}(D)$, $\Delta \mu =0$ (and $\phi =1/2$),
there is no remnant of the prewetting phenomena left, and the phase diagram
both in the ($T,\Delta \mu $) plane as well as in the ($T,\phi $) plane
looks qualitatively exactly like in the bulk three-dimensional case. Of
course, we do expect a flatter shape near the critical point due to the
occurence of the two-dimensional Ising exponents \{$\phi _{\text{coex}%
}-1/2\propto (1-T/T_{c}(D))^{1/8}$ rather than $\phi _{\text{coex}%
}-1/2\propto (1-T/T_{\text{cb}})^{\beta }$ with $\beta \approx 0.325$ in the
bulk \cite{1,2,3,4,5,9}\}. But the situation differs very much for $D>D_{t}$
(middle and right part of Fig.~1). In the semi-grand-canonical ensemble ($%
\Delta \mu $ fixed), one experiences a single first order transition for $%
-\Delta \mu _{c}(D)<\Delta \mu <\Delta \mu _{c}(D)$ where $\pm \Delta \mu
_{c}(D)$ is the chemical potential reached for $T=T_{c}(D)$ along the
remnants of the prewetting lines. In the canonical ensemble $(\phi $ fixed),
we encounter a single first-order phase transition only if $\phi =\phi _{%
\text{trip}}=\frac{1}{2}$, which corresponds to $\Delta \mu =0$, while for $%
\phi _{1}<\phi _{\text{trip}}$ or for $\phi _{\text{trip}}<\phi _{2}=1-\phi
_{1}$ one encounters two successive first-order transitions when one lowers
the temperature. Only if one chooses $\phi =\phi _{c1}$, (the critical
concentration corresponding to the critical point $T_{c1}$) or $\phi =\phi
_{c2}=1-\phi _{c1}$ (corresponding to $T_{c2}$) one still encounters a
(two-dimensional) critical behavior of phase separation, $\phi _{\text{coex}%
}-\phi _{c1}\propto (1-T/T_{c1})^{1/8}$, and similarly $\phi _{\text{coex}%
}-\phi _{c2}\propto (1-T/T_{c2})^{1/8}$. While $\phi _{1}$ as well as $\phi
_{2}$ merge at $\phi _{\text{trip}}=1/2$ as $D\rightarrow D_{t}$, $\phi _{1}$
and $\phi _{2}$ move outwards towards the prewetting critical concentration
when $D\rightarrow \infty $. In this limit, the width of the two-phase
coexistence regions for $T>T_{\text{trip}}$ must shrink and ultimately
vanish, since the difference in concentration on both sides of the
pseudo-prewetting first order lines with respect to the average
concentration $\phi $ of the film is an effect of order $1/D$, and therefore
\ for $D\rightarrow \infty $ the prewetting transitions are lines in the $%
(T,\phi )$ phase diagram and not split up in two-phase coexistence regions.

From this description it already is clear that the approach to the phase
diagram of the bulk $(D=\infty )$ as $D\rightarrow \infty $ is very
nonuniform: for any finite $D$ the bulk transition is still rounded, and
(for $\phi _{1}<\phi _{\text{trip}}$ or $\phi _{\text{trip}}>\phi _{1}$) the
first transition is a lateral phase separation corresponding to the
prewetting transition and the second transition is another lateral phase
separation at $T_{\text{trip}}$ (Fig.~2). For $D\rightarrow \infty $ the
phase diagram, hence, contains prewetting lines (as in the left part of
Fig.~1) and a horizontal line at $T_{w}$ (to which the triple line in the
middle part of Fig.~1 has converged). Of course, the pictures explaining the
various phases in Fig.~2 are highly schematic, and in reality one expects
the interface to turn around rather smoothly and avoid the 90$%
{{}^\circ}%
$ kinks. Such smooth interfaces (which also have an intrinsic width which
need not be negligibly small in comparison with $D$) have in fact been
observed in two-phase coexistence states associated with capillary
condensation \cite{10}.

\section{A quantitative example: a confined polymer mixture}

Thin polymeric films confined between walls may have interesting
applications and can also be studied conveniently by a variety of
experimental tools \cite{22}. In fact, the ''soft mode'' --
phase \cite{14} with a single fluctuating interface in the middle of the
film (as shown in the upper part of Fig.~2) has been experimentally observed 
\cite{23}, and we consider it likely that by fine-tuning of experimental
control parameters it should also be possible to observe some of the
transitions predicted in Figs.~1, 2. In fact, for polymers one need not have
special interactions to get a first-order wetting transition as in the Ising
model \cite{17}, rather one finds always first order wetting behavior except
close to the critical point of the bulk \cite{10}.

As in our previous work \cite{10,18} we consider a situation where the
wetting transition temperature $T_{w}$ lies in the strong segregation
regime, for which the self-consistent field theory is accurate. The
technical aspects of this approach are explained in detail elsewhere \cite
{10,21,24}. Fig.~3 shows that for a typical choice of parameters indeed a
phase diagram of the type of Fig.~1, right part, is reproduced. Note that
the self-consistent field theory near the critical points $T_{c1}(D)$, $%
T_{c2}(D)$ implies mean-field like behavior, $\phi _{\text{coex}}-\phi
_{c1}\propto (1-T/T_{c1}(D))^{1/2}$, rather than yielding the expected
two-dimensional Ising exponent \cite{10}. But for large molecular weight
this Ising like critical region is expected to be rather narrow \cite{22},
and thus we consider Fig.~3 as a useful hint for the phase diagram to be
searched for in the experiments.

\section{Conclusions}

In this paper we have considered the problem of phase-separating binary
mixtures confined between ''competing walls'' and have shown by qualitative
considerations (Fig.1.) and self--consistent field calculations (Fig.3.) 
that the phase
diagram has either critical points and first order regions coexisting at a
triple line or a single critical point resulting from the merging of these
two critical points at the tricritical thickness $D_{t}$. In previous work
treating the case $D>D_{t}$, only the case $\Delta \mu =0$ in the
semi-grandcanonical ensemble was studied \cite{17,18}, which in the $(T,\phi
)$ plane means that one cools the system at $\phi =\phi _{\text{trip}}=0.5$
and then a single first order transition (Fig.~2, left part) occurs: thus
the existence of the two critical points was not previously discussed.

Of course, in reality one will have to abandon the special symmetry
assumptions used in Figs.~1-3, allowing for asymmetric mixtures, differences
in strength of the wall forces, etc, and thus the space of parameters to be
considered gets much enlarged. However, as long as one works in the subspace
where the wetting transition temperatures $T_{w}$ of both walls are the
same, the phase diagrams still should have the topology of Fig.~1, only the
mirror symmetry around $\Delta \mu =\Delta \mu _{\text{coex}}(T)$ or $\phi
=\phi _{\text{cb}}$ is lost, and thus in general $\phi _{\text{trip}}$ will
differ from $\phi _{\text{cb}}$. Also $T_{c1}$ and $T_{c2}$ will differ. Of
course, in the most general case one must allow also for $T_{w1}\neq T_{w2}$%
, different wetting transition temperatures of both walls. One can also
consider first order wetting at one wall and second order wetting at the
other. A description of the phase diagrams for these more complicated cases
is a challenging task for future work.

\underline{Acknowledgements:} This work was partially supported by the DFG
under grant N$^{\text{o}}$ Bi314/17 and by the Volkswagenstiftung under
grant N$^{\text{o}}$ I/74168.

\begin{figure}[htbp]
    \begin{minipage}[t]{160mm}%
       \mbox{
        \setlength{\epsfxsize}{16cm}
        \epsffile{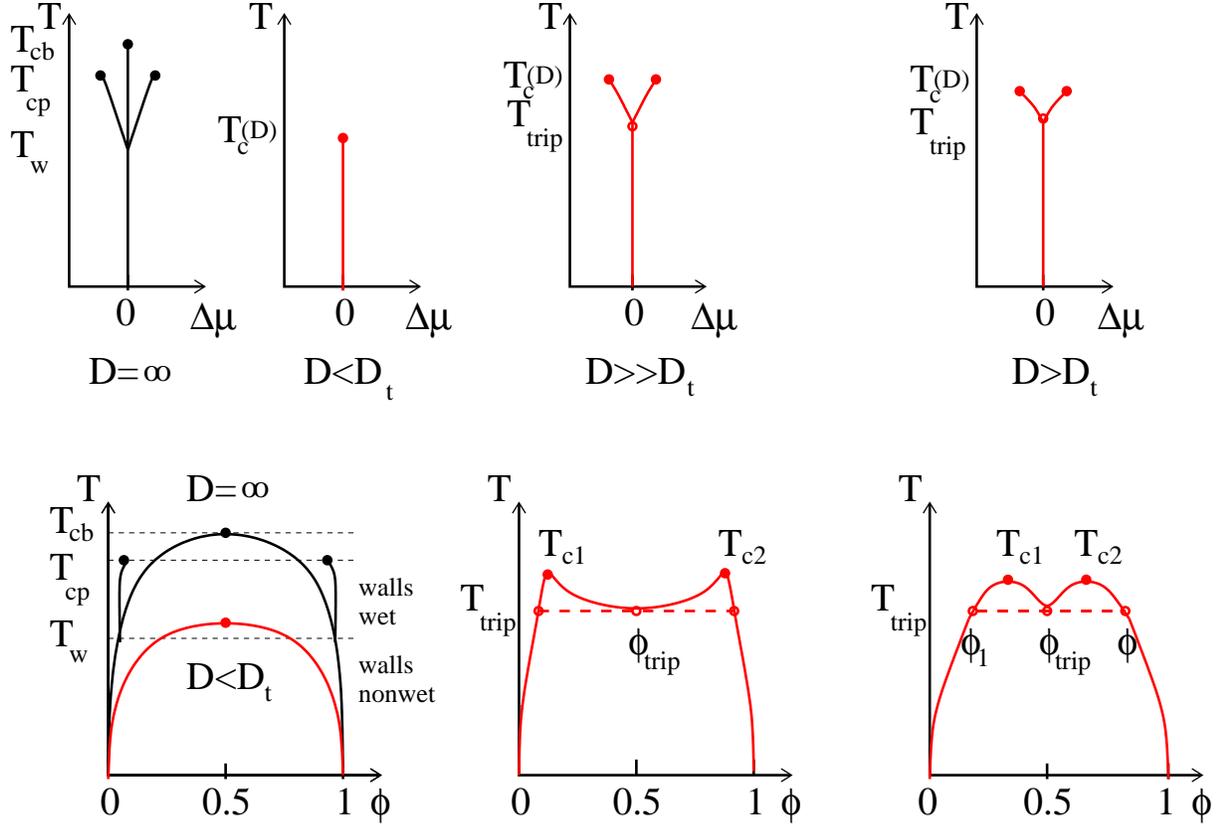}
       }\\
    \end{minipage}%
    \hfill%
    \begin{minipage}[b]{160mm}%
    \vspace*{1cm}
\caption{Qualitative phase diagrams of a bulk system (thickness $D=\infty $)
confined by two walls at which competing ''fields'' act, as well as
corresponding phase diagrams of thin films of various thicknesses $D=D<D_{t}$
(left part), $D>>D_{t}$ (middle part) and $D>D_{t}$ (right part). Upper part
of the panel presents the phase diagrams in the space of two intensive
variables, lower part chooses as abscissa instead the concentration $\protect%
\phi $, a density of an extensive variable. Characteristic temperatures
shown are the bulk critical (T$_{cb})$, wetting ($T_{w}$) and prewetting
critical ($T_{cp})$ temperatures, while in the thin film critical
temperatures $T_{c}(D)$ occur at $\Delta \protect\mu =0$ and $\protect\phi =%
\protect\phi _{cb}=1/2$ only for $D<D_{t}$, while for $D>D_{t}$ one has for $%
\Delta \protect\mu =0$ and $\protect\phi _{trip}=1/2$ rather a triple point
of three phase coexistence. However, two critical points $%
T_{c1}=T_{c2}=T_{c}(D)$ still occur, but at concentrations $\protect\phi %
_{1c},\protect\phi _{2c}$ that move towards the concentrations of the
prewetting critical points as $D\rightarrow \infty $. For further
explanations cf. text.}
\label{Fig.1}
    \end{minipage}%
\end{figure}

\begin{figure}[htbp]
    \begin{minipage}[t]{160mm}%
       \mbox{
        \setlength{\epsfxsize}{10cm}
        \epsffile{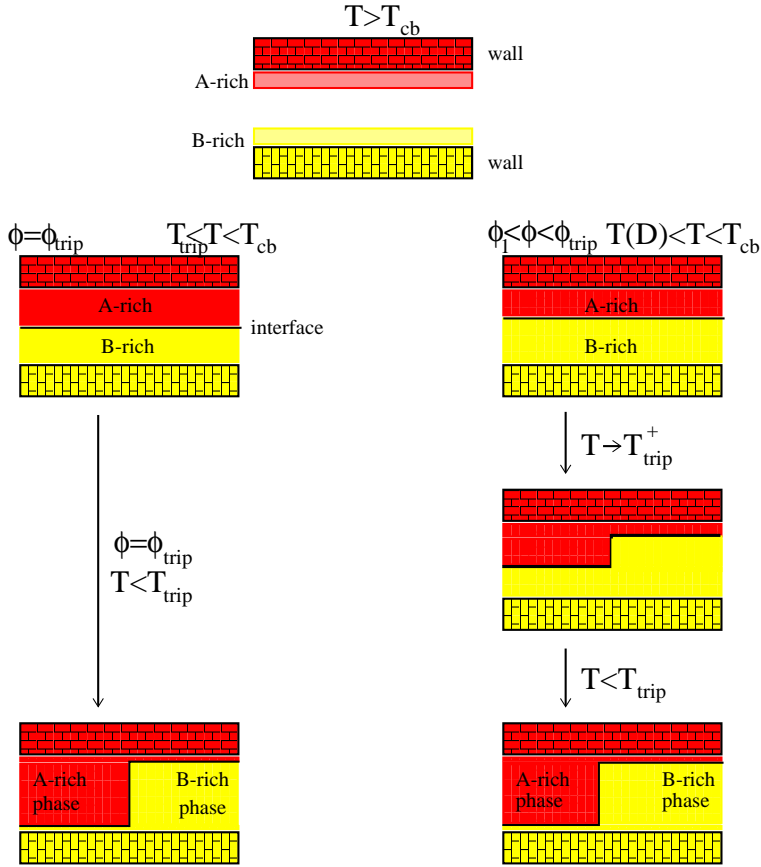}
       }
    \end{minipage}%
    \hfill%
    \begin{minipage}[b]{160mm}%
    \vspace*{1cm}
\caption{Qualitative explanation of the different phases that occur in a
binary mixture confined between competing walls. Already at $T>T_{cb}$ there
are A-rich and B-rich enrichment layers at the respective walls, having a
finite thickness $\protect\xi _{b}$ (the bulk correlation length of
concentration fluctuations), so for $D>>2\protect\xi _{b}$ the system is
still disordered in the center, while in the region of temperatures where $D$
and $\protect\xi _{b}$ are comparable a rounded transition to a already
structure with one A-B interface at $T \stackrel{<}{~} T_{cb}$ occurs. For $\protect%
\phi =\protect\phi _{trip}$ the position of this interface is just in the
middle of the thin film, while for $\protect\phi _{1}<\protect\phi <\protect%
\phi _{trip}$ the location of the interface reflects the asymmetry of
composition. For $\protect\phi =\protect\phi _{trip}$ one encounters a
single transition at $T=T_{trip}$, where the localization of the interface
at the walls requires lateral phase separation between A-rich and B-rich
phases of equal amounts. For $\protect\phi _{1}<\protect\phi <\protect\phi
_{trip}$ one encounters two lateral phase separations: the first one is a
coexistence between the phase with delocalized interface (in the center of
the film as $T\rightarrow T_{trip}^{+}$) with a phase where the interface is
localized at the upper wall (the B-rich phase). In a second transition at $%
T<T_{trip}$ the phase with delocalized interface disappears in favor of the
phase with the interface bound to the lower wall (the A-rich phase). Note
that the amount of this phase must be less, to comply with the lever rule.}
\label{Fig.2}
    \end{minipage}%
\end{figure}

\begin{figure}[htbp]
    \begin{minipage}[t]{160mm}%
       \mbox{
        \setlength{\epsfxsize}{8cm}
        \epsffile{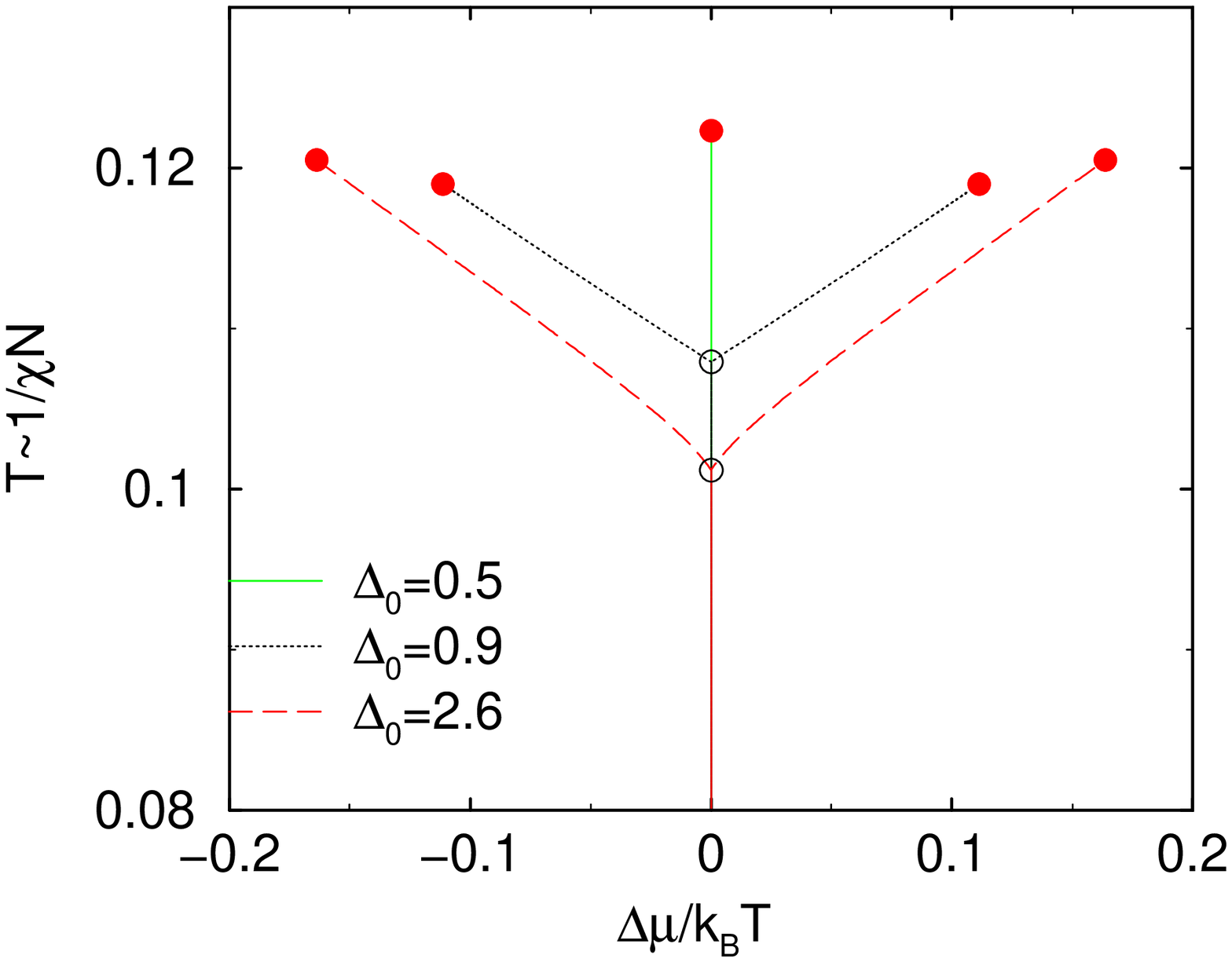}
        \setlength{\epsfxsize}{8cm}
        \epsffile{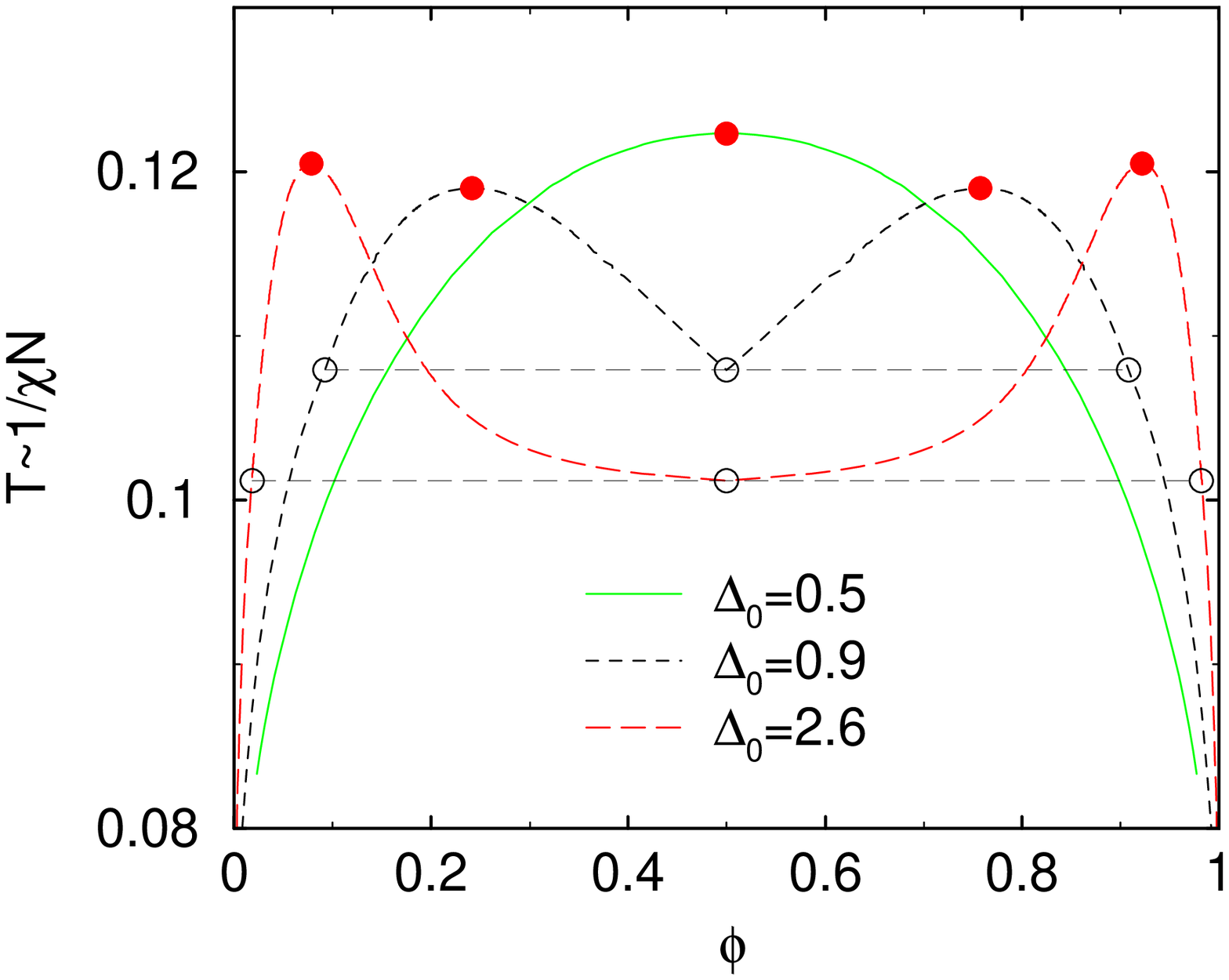}
       }
    \end{minipage}%
    \hfill%
    \begin{minipage}[b]{160mm}%
    \vspace*{1cm}
\caption{Phase diagram of a symmetric polymer mixture in a thin film with
antisymmetric boundary fields calculated in the framework of the
self-consistent field theory for Gaussian chain molecules. Three film
thicknesses $D=0.5R_{e},0.9R_{e}$ and $2.6R_{e}$ are shown ($R_{e}$:
end-to-end distance of the molecules). The smallest film thickness
corresponds to the situation $D<D_{t}$ while films of thickness $D=0.9R_{e}$
and $2.6R_{e}$ exhibit two critical points. Panel (a) displays the phase
diagrams in the $\ $($T\sim \frac{1}{\protect\chi N},\Delta \protect\mu /kT$%
) plane, while (b) shows the results as a function of temperature $\frac{1}{%
\protect\chi N}$ and composition $\protect\phi $. Note that the critical
temperature of the bulk is at \protect{$1/\chi N =0.5$}.}
\label{Fig.3}
    \end{minipage}%
\end{figure}

\end{document}